\documentclass[11pt]{article}

\usepackage{amssymb}
\usepackage{amsmath}
\usepackage{nccmath}
\usepackage{graphicx}

\usepackage{color}

\newcommand{\di}{\partial}

\newcommand{\Slash}[1]{{\ooalign{\hfil/\hfil\crcr$#1$}}}

\newcommand{\dx}{{d^4}\kern-.07cm{x}}

\begin{document}

\begin{titlepage}
\begin{center}

\vspace*{10mm}

{\LARGE\bf
Constraints on gauge-Higgs unification models \\[0.3cm] at the LHC}

\vspace*{20mm}

{\large
Noriaki Kitazawa\footnote[1]{kitazawa@phys.se.tmu.ac.jp}
 and Yuki Sakai\footnote[2]{sakai-yuki1@ed.tmu.ac.jp}
}
\vspace{6mm}  

{\it
Department of Physics, Tokyo Metropolitan University,\\
Hachioji, Tokyo 192-0397, Japan\\
}

\vspace*{15mm}

\begin{abstract}
We examine the possibility of observing
 the Kaluza-Klein gluons in gauge-Higgs unification models at the LHC with the energy $\sqrt{s}=14$ TeV.
We consider a benchmark model with the gauge symmetry SU$(3)_C \times$ SU$(3)_W$ in five-dimensional space-time,
 where SU$(3)_C$ is the gauge symmetry of the strong interaction
 and SU$(3)_W$ is that for the electroweak interaction and a Higgs doublet field.
It is natural in general
 to introduce SU$(3)_C$ gauge symmetry in five-dimensional space-time as well as SU$(3)_W$ gauge symmetry
 in gauge-Higgs unification models.
Since the fifth dimension is compactified to $S^1/{\bf Z}_2$ orbifold,
 there are Kaluza-Klein modes of gluons in low-energy effective theory in four-dimensional space-time.
We investigate the resonance contribution of the first Kaluza-Klein gluon
 to dijet invariant mass distribution at the LHC,
 and provide signal-to-noise ratios in various cases of
 Kaluza-Klein gluon masses and kinematical cuts.
Although the results are given in a specific benchmark model,
 we discuss their application to general gauge-Higgs unification models with Kaluza-Klein gluons. 
Gauge-Higgs unification models can be verified or constrained
 through the physics of the strong interaction,
 though they are proposed to solve the naturalness problem in electroweak symmetry breaking.
\end{abstract}

\end{center}
\end{titlepage}

\section{Introduction}

Gauge-Higgs unification (GHU) model
 is a candidate of the physics beyond the Standard Model (SM)
 in higher dimensional space-time with compactified extra dimensions.
It is one of the models with large extra dimensions  \cite{Antoniadis:1990ew,ArkaniHamed:1998rs,ArkaniHamed:1998nn}.
In general GHU models
 there are infinite number of Kaluza-Klein (KK) modes of the fields in higher dimensional space-time,
 which correspond to infinite number of discretized momenta in extra dimensions.
The KK zero modes,
 which correspond to the states with zero momenta in extra dimensions,
 are massless in low-energy effective theory in four-dimensional space-time.
The lowest KK non-zero modes, ``the first KK modes'', have masses of the order of compactification scales
 which are the inverse of the sizes of the extra dimensions.
The KK zero mode of a gauge field is decomposed into two parts:
 the components of four-dimensional space-time consist a gauge field
 and the components of extra dimensions are scalar fields in four-dimensional space-time.
The explicit gauge symmetry breaking at the orbifold fixed points
 in the orbifold compactification of extra dimensions \cite{Kawamura:1999nj}
 makes those scalar fields possible to be Higgs doublet fields for electroweak symmetry breaking.

The basic idea of GHU model were introduced in
 \cite{Manton:1979kb,Fairlie:1979at,Hosotani:1983xw,Hosotani:1983vn,Hosotani:1988bm}.
The finiteness of the quantum correction to the Higgs mass
 is expected by the higher dimensional gauge invariance,
 since the Higgs doublet field originates from higher dimensional gauge field
 \cite{Hatanaka:1998yp,Antoniadis:2001cv}.
The basic idea for realistic mass hierarchy among quarks and leptons were proposed in
 \cite{Hall:2001zb,Csaki:2003dt,Burdman:2002se,Scrucca:2003ra}.
It is a non-trivial problem,
 because the Higgs doublet field universally interacts with all the fermions
 due to the restriction of gauge principle.
In \cite{Hall:2001zb,Csaki:2003dt,Scrucca:2003ra}
 quarks and leptons are assumed to be localized at orbifold fixed points,
 and the mixings with heavy extra fermions living in the whole space-time, ``bulk'',
 generate their Yukawa couplings with Higgs doublet field.
In \cite{Burdman:2002se}
 quarks and leptons are introduced as bulk fermions,
 and their non-trivial distributions in extra dimensions due to their bulk masses
 generate Yukawa couplings with Higgs doublet field.

In contrast to extensive studies on electroweak sector,
 less attention has been payed on the strong interaction or QCD sector.
In view of the above idea of GHU model it is natural to assume that
 the gauge symmetry of the strong interaction is also the gauge symmetry in higher dimensional space-time.
Therefore, KK gluons naturally exist in GHU models in general.
Searching for the signals of KK gluons gives constraints to the models,
 because the masses of KK gluons are related to the sizes of compactified extra dimensions.
The sizes of compact extra dimensions
 are directly connected to the energy scale of electroweak symmetry breaking in GHU models
 \cite{Hosotani:1983xw,Hosotani:1983vn,Hosotani:1988bm},
 and too small sizes of compact extra dimensions,
 or too heavy KK gluons, contradict the observed electroweak scale.
Although the constraints to the sizes of compact extra dimensions through the KK modes of electroweak gauge bosons
 have been discussed in \cite{Maru:2007xn,Maru:2013ooa,Maru:2013bja}, for example,
 GHU models can be tested in a complementary way from the physics of the strong interaction.
Notice that GHU models without KK gluon may be possible,
 and we do not discuss the constraints to these class of models in this paper.

There have already been many investigations into the physics of KK gluons at the LHC,
 in case of flat compact spaces \cite{Nath:1999mw,Dicus:2000hm}
 and in case of warped extra dimensions \cite{Agashe:2006hk,Lillie:2007ve,Haba:2012bc}.
Systematic general studies of the massive color-octet vector boson productions
 have already been given in \cite{Chivukula:2011ng,Chivukula:2013xla}.
In this paper we specially concentrate on the physics of KK gluons in GHU models.

We discuss the constraints to the mass of KK gluon
 in a simple benchmark GHU model in five-dimensional space-time with an orbifold compactification.
The gauge symmetry of the strong interaction is introduced in the bulk as well as the electroweak gauge symmetry.
This benchmark model is used
 as a starting point to discuss general nature of the KK gluon in GHU models,
 and the pursuit of realistic models is beyond the scope of this paper.
We investigate
 the signal of the first KK gluon state in dijet process, $pp \rightarrow jj$, at the parton level.
We further discuss how the results apply to more general GHU models with KK gluons.

This paper is organized as follows.
In the next section,
 we introduce a benchmark GHU model with SU$(3)_C \times$SU$(3)_W$ gauge symmetry in five-dimensional space-time.
The fifth dimension is compactified in $S^1/{\bf Z}_2$ orbifold.
We investigate in detail the couplings between quarks and KK gluons
 which depend on the Yukawa couplings of the quarks under the scenario in \cite{Burdman:2002se}.
In section \ref{sec:result},
 we discuss the contribution of the KK gluon to dijet invariant mass distributions
 at the LHC with the energy $\sqrt{s}=14$ TeV.
The signal-to-noise ratios in various cases of KK gluon masses and kinematical cuts are given,
 and the constraints on the benchmark GHU model is discussed.
The arguments to translate the constraints to general GHU models with KK gluons are given.
In section \ref{sec:summary}, we summarize our results.

\section{Kaluza-Klein gluons in gauge-Higgs unification models}
\label{sec:model}
In this section we introduce a simplest five-dimensional GHU model
 in which the gauge symmetry SU$(3)_C \times$SU$(3)_W$ is imposed in five-dimensional space-time.
The gauge symmetry SU$(3)_C$
 is related to the gauge symmetry of the strong interaction in four-dimensional space-time.
We should extend the electroweak gauge symmetry, because the Higgs field is electroweak doublet.
We consider SU$(3)_W$ gauge symmetry
 as a minimal extension containing the electroweak gauge bosons and a Higgs doublet.
The compactification of an extra dimension in $S^1/{\bf Z}_2$ orbifold realizes the chiral theory.
The coordinate of the extra dimension, $y$, takes the value in the interval $[0,L]$.
Since SU$(3)_C$ gauge symmetry,
 which results the gauge symmetry of the strong interaction in four-dimensional space-time,
 is imposed in five-dimensional space-time,
 gluon field is expanded in KK modes.

The action and Lagrangian density of the benchmark model
 with the fifth dimension compactified on $S^1/{\bf Z}_2$ orbifold is
\begin{eqnarray}
	\label{eq:toylagrangian}
	S= {\int}\kern-.1cm{\dx} dy \, {\cal L}, \quad
	{\cal L} = -\frac{1}{4}G^a_{MN}G^{aMN} - \frac{1}{4}F^a_{MN}F^{aMN}
	           + \bar{\Psi}i\Slash{D} \Psi -\epsilon(y)M\bar{\Psi}\Psi,
\end{eqnarray}
 where indices of capital letters run over 0 to 4, and $D$ stands for covariant derivative, 
\begin{eqnarray}
	\Slash{D} \equiv \Gamma^MD_M, \quad \Gamma^M = ( \gamma^\mu,i\gamma^5).
\end{eqnarray}
Here, $G$ and $F$ denote field strengths corresponding to SU$(3)_C$ and SU$(3)_W$, respectively.
\begin{eqnarray}
	G^a_{MN} = \di_M G^a_N - \di_N G^a_M - g_5f^{abc}G^b_M G^c_N, \quad 
	F^a_{MN} = \di_M A^a_N - \di_N A^a_M - g'_5f^{abc}A^b_M A^c_N. 
\end{eqnarray}
The indices of small letters stand for adjoint indices of SU$(3)$.
The function $\epsilon(y)$ is the sign function and $M$ is the bulk mass of the fermion.
Though the function can be any anti-symmetric functions in general,
 we choose this simple form of the bulk mass term.
In this paper we adopt the mechanism for fermion mass hierarchy of \cite{Burdman:2002se},
 but note that detailed structure to obtain realistic masses and mixings of quarks and leptons
 does not affect our final results.

The invariance of the action under
 the translation $y$ to $y+2L$ (corresponding to $S^1$ compactification)
 and the parity $y$ to $-y$ (corresponding to ${\bf Z}_2$ identification)
 sets the boundary conditions to the fields.
\begin{eqnarray}
	\label{eq:BCfermion}
	&&\Psi(x,y) = \Psi(x,y+2L) = P\gamma_5 \Psi(x,-y), \\
	\label{eq:BCgauge1}
	&&A_\mu(x,y) = A_\mu(x,y+2L) = PA_\mu(x,-y)P^\dagger, \\
	\label{eq:BCgauge2}
	&&A_4(x,y) = A_4(x,y+2L) = -PA_4(x,-y)P^\dagger.
\end{eqnarray}
Here, $P$ is a certain unitary matrix which belongs to SU$(3)_W$.
The boundary condition Eq.(\ref{eq:BCfermion}) realizes chiral fermions
 in low-energy effective theory in four-dimensional space-time.
If we choose $P=\mathrm{diag}(-1,-1,1)$,
 the gauge symmetry is broken at the orbifold fixed points from SU$(3)_W$ to SU$(2)\times$U$(1)$
 by the boundary conditions Eqs.(\ref{eq:BCgauge1}) and (\ref{eq:BCgauge2}).
The only components of the fields with even ${\bf Z}_2$ parity can have the massless KK zero modes.
The electroweak gauge fields and Higgs doublet field
 arise from the KK zero modes of $A_\mu$ and $A_4$, respectively.
On the other hand, the boundary conditions for SU$(3)_C$ gauge fields are set to be
\begin{eqnarray}
	\label{eq:BCgauge3}
	&&G_\mu(x,y) = G_\mu(x,y+2L) = G_\mu(x,-y), \\
	\label{eq:BCgauge4}
	&&G_4(x,y) = G_4(x,y+2L) = -G_4(x,-y),
\end{eqnarray}
 so that unbroken SU$(3)_C$ gauge symmetry of the strong interaction is realized
 in four-dimensional space-time.
These boundary conditions show that
 all the fields are on a circle, $y \in [0,2L)$ with the point $y=2L$ identified with $y=0$,
 and the boundaries $y=0,L$ are the fixed points of ${\bf Z}_2$ orbifold.
We decompose the fermion $\Psi$
 into two chiral components in four-dimensional space-time, $\Psi_\pm$,
 where
\begin{eqnarray}
	\label{eq:decomposition}
	\Psi = \Psi_+ + \Psi_-, \qquad \gamma_5\Psi_\pm = \pm\Psi_\pm.
\end{eqnarray}
The configurations of the fermion field $\Psi_+$
 and gauge fields with even parity are symmetric at the fixed points,
 and the configurations of the fermion field $\Psi_-$
 and gauge fields with odd parity are antisymmetric at the fixed points.

The KK mode expansions of the gauge fields are
\begin{eqnarray}
	A^+_M(x,y) &=&
	 \frac{1}{\sqrt{2L}}
	  \left[ A^{+(0)}_M(x) + \sqrt{2} \sum^\infty_{n=1} A^{+(n)}_M(x) \cos\left( \frac{n\pi}{L} y\right) \right], \\
	A^-_M(x,y) &=&
	 \frac{1}{\sqrt{L}}
	  \sum^\infty_{n=1} A^{-(n)}_M(x) \sin\left( \frac{n\pi}{L} y\right), \\
	G^+_\mu(x,y) &=&
	 \frac{1}{\sqrt{2L}}
	  \left[ G^{+(0)}_\mu(x) + \sqrt{2} \sum^\infty_{n=1} G^{+(n)}_\mu(x) \cos\left( \frac{n\pi}{L} y\right) \right], \\
	G^-_4(x,y) &=&
	 \frac{1}{\sqrt{L}}
	  \sum^\infty_{n=1} G^{-(n)}_4(x) \sin\left( \frac{n\pi}{L} y\right),
\end{eqnarray}
 where $+$ or $-$ denote the parity assignment.
The mass of the field of $n$-th KK mode is given as $m_n = n\pi/L$
 which depends on the size of the compact extra dimension $L$.
The fermion fields are expanded by an orthonormal set of mode functions
 which depends on the form of the bulk mass term.
\begin{eqnarray}
	\Psi_\pm (x,y) = \sum_{n=0} \Psi^{(n)}_\pm(x) f^{(n)}_\pm (y).
\end{eqnarray}
The zero mode functions satisfy
\begin{eqnarray}
	[ \mp \di_4 + M \epsilon(y) ] f^{(0)} _\pm (y) = 0
\end{eqnarray}
 under the present simple form of the bulk mass,
 and we obtain the zero mode functions as
\begin{eqnarray}
	f^{(0)}_-(y) = \sqrt{\frac{M}{1-e^{-2 ML}}}e^{-M|y|}, \qquad f^{(0)}_+(y) = \sqrt{\frac{M}{e^{2 ML}-1}}e^{M|y|}.
\end{eqnarray}
These functions can be understood as
 the wave functions of the zero mode fermions in the extra dimension.

%
%
%
%
\begin{figure}[t]
\begin{center}
    \includegraphics[width=15.0cm]{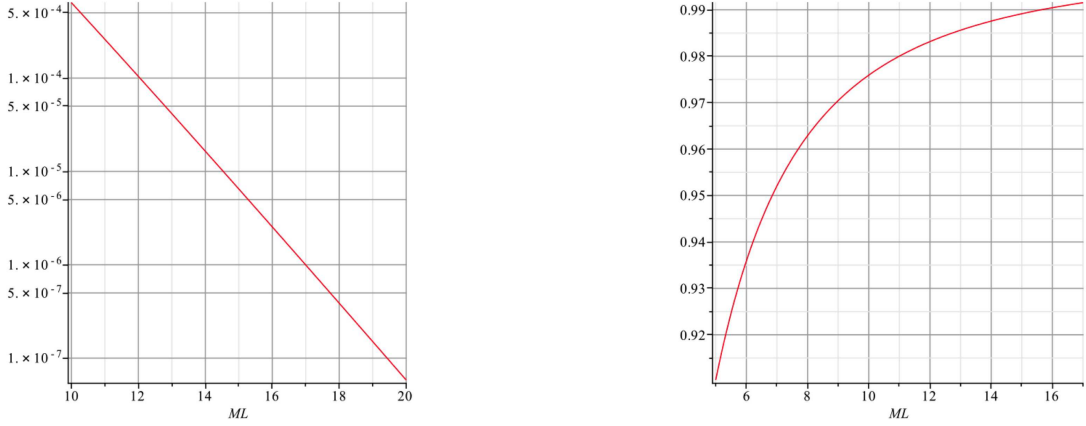}
    \vspace{0cm}
 \caption{
The dependence of the coefficient of $\sqrt{2} g_2$ in Eq.(\ref{eq:Yukawacouplingcoefficient}) on $ML$ (Left panel).
The dependence of the coefficient of $g_5/\sqrt{L}$ in Eq.(\ref{eq:KKonemodecouplings}) on $ML$ (Right panel).
 }
\label{fig:couplings}
\end{center}
\end{figure}
The couplings in four-dimensional effective theory are given by the integrals over the extra dimension.
The couplings between zero mode fermion fields and gluon field are obtained as
\begin{eqnarray}
	g_L = g_5\int^{L}_{-L} \frac{1}{\sqrt{2L}} | f^{(0)}_- (y) |^2 dy,
	\qquad
	g_R = g_5\int^{L}_{-L} \frac{1}{\sqrt{2L}} | f^{(0)}_+ (y) |^2 dy,
\end{eqnarray}
 where $g_L$ and $g_R$ correspond to the gluon couplings with left-handed and right-handed fermions, respectively.
These coupling constants should coincide with each other, because the strong interaction is vector-like in the SM.
In fact performing the integrals yields the relation of $g_5$ to the strong coupling constant $g_s$.
\begin{eqnarray}
	g_L = g_R = \frac{g_5}{\sqrt{2 L}} \equiv g_s.
\end{eqnarray}
Through the same procedure we obtain the relation of $g'_5$ to weak coupling, $g_2$, in the SM.
\begin{eqnarray}
	g_2 = \frac{g'_5}{\sqrt{2 L}}.
\end{eqnarray}
The couplings between the first KK gluon, $G_\mu^{+(1)}$, and zero mode fermions are obtained as
\begin{eqnarray}
\label{eq:KKonemodecouplings}
	g^{(1)}_L = -g^{(1)}_R
	 = \frac{4M^2 L^2}{4M^2 L^2 + \pi^2} \frac{1+ e^{-2 ML}}{1 - e^{-2 ML}} \times \frac{g_5}{\sqrt{L}}\equiv g^{(1)}_s.
\end{eqnarray}
In large $ML$ limit the coefficient of $g_5/\sqrt{L}$ becomes to 1, and we obtain a simple relation
\begin{eqnarray}
	\label{eq:KKgluoncouplings}
	g^{(1)}_s \simeq \sqrt 2 g_s.
\end{eqnarray}
Note that the first KK gluon field behaves as an axial vector field, like axigluon \cite{Frampton:1987dn}.
This is the result of choosing a simple form of the bulk mass term in Eq.(\ref{eq:toylagrangian}).
In general the coefficient function $\epsilon(y)$ may be any anti-symmetric functions,
 and the first KK gluon field is not necessarily an axial vector field.
In the following we consider this simplest case for a while, and later we will discuss what happens in general cases.

We show the validity of assuming large $ML$ in the calculations of the amplitudes of the processes at the LHC.
The Yukawa coupling between Higgs doublet and fermions is given by
\begin{eqnarray}
	\label{eq:Yukawacouplingcoefficient}
	\int^{L}_{-L} \sqrt{ \frac{M}{1-e^{-2 ML}}} e^{-M|y|} \frac{g'_5}{\sqrt{2L}} \sqrt{\frac{M}{e^{2 ML} -1}} e^{M|y|} dy
	  = \frac{ML}{\sqrt{\cosh(2 ML) - 1}} \times \sqrt{2}g_2.
\end{eqnarray}
Since we consider the physics at the LHC, the fermion should be up or down quark,
 and their Yukawa coupling is ${\cal O}(10^{-5})$ times $\sqrt{2}g_2$.
The left panel of Fig.~\ref{fig:couplings}
 shows the shape of the function in front of $\sqrt{2}g_2$ in Eq.(\ref{eq:Yukawacouplingcoefficient}).
We can read $ML = 14 \sim 15$ for small Yukawa couplings of up and down quarks.
The right panel of Fig.~\ref{fig:couplings}
 shows the shape of the function in front of $g_5/\sqrt{L}$ in Eq.(\ref{eq:KKonemodecouplings}).
The value of $ML = 14 \sim 15$ results the value of the function to be about 0.99.
Therefore, the relation (\ref{eq:KKgluoncouplings}) can be used for our aim.
Of course this result strongly depends on the present simple form of the bulk mass,
 and the value $g^{(1)}_s = \sqrt 2 g_s$ can be translated to be the maximal value of $g^{(1)}_s$.
It is worth to mention that
 in the models with quarks localized at orbifold fixed points,
 it has been shown that the couplings of KK gluon with massless quarks are also $\sqrt{2}g_s$ \cite{Dicus:2000hm}.
We will take into account the model dependence by dealing $g^{(1)}_s$ with a parameter less than $\sqrt{2}g_s$.

\section{Search for the KK gluon in dijet invariant mass distribution}
\label{sec:result}

In order to search for the KK gluon in the process of $pp \to jj$ at the LHC experiments,
 we first consider the scatterings of two partons into two parton final states.
Next, we take into account the parton distribution functions of proton
 and obtain the cross section of the process $pp \to jj$
 under the assumption that two outgoing partons generate two jets.
Then, we calculate the dijet invariant mass distributions
 with various values of the mass of KK gluon and various kinematical cuts.

The squared amplitude of the process of two identical partons,
 $q = u, d$ or their anti-particles, scattered by one KK gluon exchanges is
\begin{eqnarray}
	\label{eq:amplitude1}
	|{\cal M}(qq \rightarrow qq)|^2 &=&
	 16\frac{(8\pi\alpha_s)^2}
	        {(\hat t - m_1^2)^2 + (m_1 \Gamma)^2}
	 (\hat s^2 + \hat u^2)
   + 16\frac{(8\pi\alpha_s)^2}
            {(\hat u - m_1^2)^2 + (m_1 \Gamma)^2}
     (\hat s^2 + \hat t^2)
\nonumber \\
	&& +
	 \frac{32}{3}
	 \frac{(8\pi\alpha_s)^2\left\{(\hat t - m_1^2)(\hat u - m_1^2) + (m_1\Gamma)^2 \right\}}
	      {\left\{ (\hat t - m_1^2)^2 + (m_1\Gamma)^2 \right\} \left\{ (\hat u - m_1^2)^2 + (m_1\Gamma)^2 \right\}}
	 \hat s^2
\nonumber \\
	&\equiv& |{\cal M}_{qq}|^2,
\end{eqnarray}
 where $\alpha_s = g_s^2/4\pi$, $\hat s$, $\hat t$ and $\hat u$ stand for Mandelstam variables of partons,
 and $m_1$ and $\Gamma$ are the mass and width of the first KK gluon, respectively.
Since the present KK gluon is axial vector, gluon fusion can not generate single KK gluon.
We will discuss the possibility of gluon fusion at the end of this section.
The squared amplitudes of the process of parton and antiparton of the same flavor scattered by one KK gluon exchanges are
\begin{eqnarray}
	\label{eq:s-channels}
	|{\cal M}(q\bar{q} \rightarrow q\bar{q})|^2 &=&
	 16\frac{(8\pi\alpha_s)^2}{(\hat s - m_1^2)^2 + (m_1 \Gamma)^2}
	  (\hat t^2 + \hat u^2)
   + 16\frac{(8\pi\alpha_s)^2}{(\hat t - m_1^2)^2 + (m_1 \Gamma)^2}
      (\hat s^2 + \hat u^2) \nonumber \\
  &&+ \frac{32}{3}\frac{(8\pi\alpha_s)^2\left\{(\hat s - m_1^2)(\hat t - m_1^2) + (m_1\Gamma)^2 \right\}}
                       {\left\{ (\hat s - m_1^2)^2 + (m_1\Gamma)^2 \right\} \left\{ (\hat t - m_1^2)^2 + (m_1\Gamma)^2 \right\}}
      \hat s^2 ,\\
	\label{eq:ubaruq}
	|{\cal M}(q\bar{q} \rightarrow q'\bar{q}')|^2 &=&
	 16 \frac{(8\pi\alpha_s)^2}{(\hat s - m_1^2)^2 + (m_1 \Gamma)^2}
	  (\hat t^2 + \hat u^2).
\end{eqnarray}
We define $|{\cal M}_{q\bar{q}}|^2$ as the sum of these two squared amplitudes
\begin{eqnarray}
	\label{eq:amplitude2}
	|{\cal M}_{q\bar{q}}|^2 &\equiv&
	 N_f \times16\frac{(8\pi\alpha_s)^2}{(\hat s - m_1^2)^2 + (m_1 \Gamma)^2}
	  (\hat t^2 + \hat u^2) 
   + 16\frac{(8\pi\alpha_s)^2}{(\hat t - m_1^2)^2 + (m_1 \Gamma)^2}
	  (\hat s^2 + \hat u^2) \nonumber \\
	&+& \frac{32}{3}
	    \frac{(8\pi\alpha_s)^2\left\{(\hat s - m_1^2)(\hat t - m_1^2) + (m_1\Gamma)^2 \right\}}
	         {\left\{ (\hat s - m_1^2)^2 + (m_1\Gamma)^2 \right\} \left\{ (\hat t - m_1^2)^2 + (m_1\Gamma)^2 \right\}}
	    \hat s^2,
\end{eqnarray}
 where we introduce $N_f \geq 2$ as the number of quarks
 which couple to the KK gluon with the maximal value of the coupling $\sqrt{2} g_s$.
In our benchmark model
 we only consider the quarks which couple to the KK gluon with the maximal coupling.
The squared amplitude of the process of two different partons scattered by one KK gluon exchanges is
\begin{eqnarray}
	\label{eq:amplitude3}
	|{\cal M}(qq' \rightarrow qq')|^2 = 
	16\frac{(8\pi\alpha_s)^2}
	       {(\hat t^2 - m_1^2)^2 + (m_1\Gamma)^2}
	  (\hat s^2 + \hat u^2)
    \equiv |{\cal M}_{qq'}|^2.
\end{eqnarray}
The decay width of the first KK gluon is given by $\Gamma = N_f \alpha_sm_1/3$
 assuming its decay to $N_f$ quark anti-quark pairs with the maximal coupling.
These squared amplitudes give the signal of KK gluon in dijet invariant mass distribution.
On the other hand,
 the ordinary QCD process by gluon exchanges produce the background in dijet invariant mass distribution.

\begin{figure}[t]
  \begin{center}
      \includegraphics[width=17.0cm]{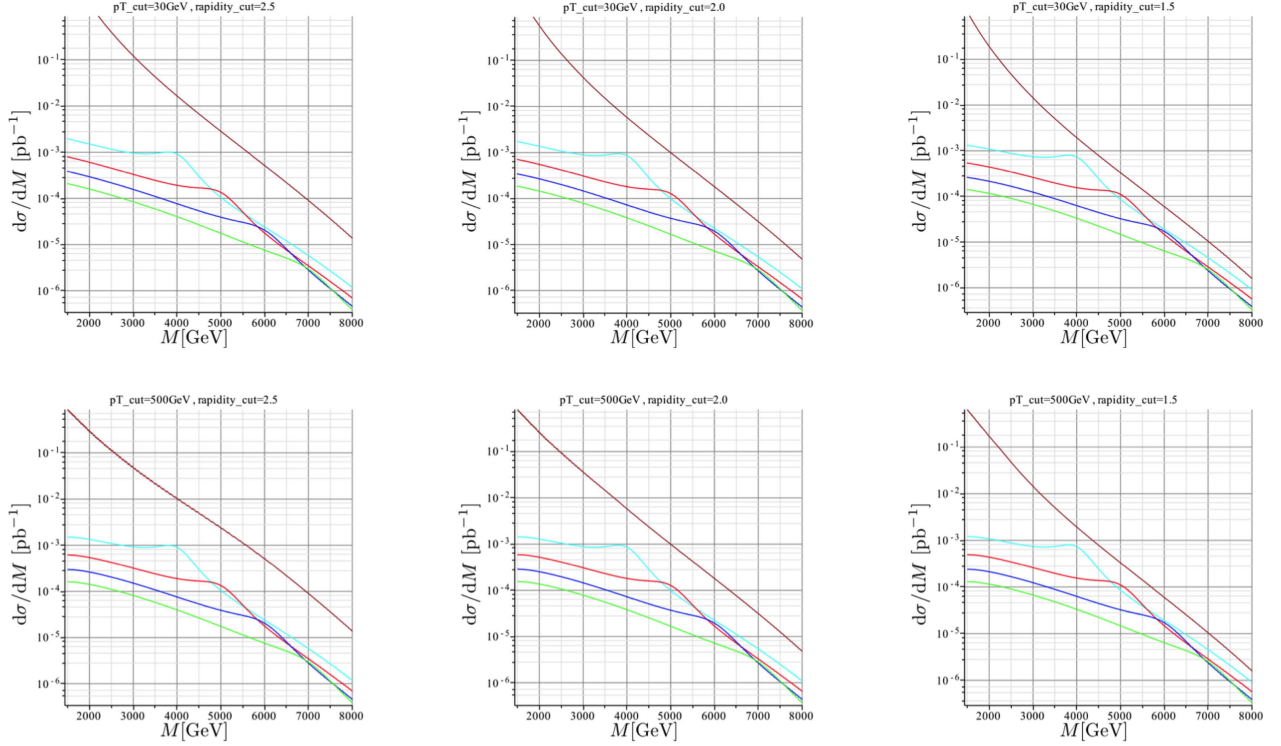}
      \hspace{0cm} 
      \vspace{0cm}
   \caption{
Dijet invariant mass distributions
 by ordinary two body QCD scattering processes (brown) and by one KK gluon exchanges
 in case with rapidity cuts $|y|\leq$ 2.5 (left column), 2.0 (middle column) and 1.5 (right column)
 with $p^{\rm min}_T=30$ GeV (upper row) and $p^{\rm min}_T=500$ GeV (lower row)
 at $\sqrt{s}=14$ TeV.
We choose $N_f=6$ in the amplitude of one KK gluon exchanges.
The lines with cyan, red, blue and green
 correspond to the KK gluon masses $m_1=4, 5, 6$ and $7$ TeV, respectively.
} 
\label{fig:distributions}
 \end{center}
\end{figure}
\begin{table}[t] 
   \begin{center}
   \begin{tabular}{c}
\begin{minipage}{0.5\hsize}
 \begin{center}
   \begin{tabular}{|c||c|c|c|} 
  \multicolumn{4}{c}{$S/\sqrt{B}$ with $\int Ldt= 25\mathrm{\ fb}^{-1}$} \\ \hline\hline
  \multicolumn{1}{|c||}{maximum of $|y|$} & 2.5 & 2.0 & 1.5  \\ \hline\hline
  \multicolumn{4}{|c|}{$p_T^{\rm min}$ = $30$ GeV}  \\ \hline
  $m_1$ = 4 TeV  & 24 & 37 & 55 \\
  $m_1$ = 5 TeV  & 8.6 & 14 & 21 \\
  $m_1$ = 6 TeV  & 3.1 & 5.0 & 7.7 \\ 
  $m_1$ = 7 TeV  & 1.1 & 1.8 & 2.7 \\ \hline\hline
    \multicolumn{4}{|c|}{$p_T^{\rm min}$ = $500$  GeV} \\ \hline
  $m_1$ = 4 TeV  & 30 & 37 & 55 \\
  $m_1$ = 5 TeV  & 9.4 & 14 & 21 \\
  $m_1$ = 6 TeV  & 3.2 & 5.0 & 7.7 \\ 
  $m_1$ = 7 TeV  & 1.1 & 1.8 & 2.7 \\ \hline
  \end{tabular}
 \end{center}
\end{minipage}
    \begin{minipage}{0.5\hsize}
 \begin{center}
   \begin{tabular}{|c||c|c|c|} 
  \multicolumn{4}{c}{$S/\sqrt{B}$ with $\int Ldt= 100\mathrm{\ fb}^{-1}$}\\ \hline\hline
  \multicolumn{1}{|c||}{maximum of $|y|$} & 2.5 & 2.0 & 1.5  \\ \hline\hline
  \multicolumn{4}{|c|}{$p_T^{\rm min}$ = $30$ GeV} \\ \hline
  $m_1$ = 4 TeV  & 47 & 75 & 110 \\
  $m_1$ = 5 TeV  & 17 & 28 & 42 \\
  $m_1$ = 6 TeV  & 6.3 & 10 & 15 \\ 
  $m_1$ = 7 TeV  & 2.2 & 3.5 & 5.5 \\ \hline\hline
    \multicolumn{4}{|c|}{$p_T^{\rm min}$ = $500$ GeV} \\ \hline
  $m_1$ = 4 TeV  & 60 & 75 & 110 \\
  $m_1$ = 5 TeV  & 19 & 28 & 42 \\
  $m_1$ = 6 TeV  & 6.3 & 10 & 15 \\ 
  $m_1$ = 7 TeV  & 2.2 & 3.5 & 5.5 \\ \hline
  \end{tabular}
 \end{center}
\end{minipage}
\end{tabular}
\end{center}
\caption{
Signal-to-noise ratios
 corresponding to various masses of KK gluon and kinematical cuts
 in case of $N_f=6$ at $\sqrt{s} = 14$ TeV
 with the integrated luminosity $\int Ldt= 25\mathrm{\ fb}^{-1}$ (Left table)
 and $\int Ldt= 100\mathrm{\ fb}^{-1}$ (Right table).
}
\label{tab:SNratio}
\end{table}

The formula of the dijet invariant mass distribution is derived in the following way.
The relation between the cross section of dijet events in two-proton scattering process
 and the cross sections of two-parton scattering processes is described as
\begin{eqnarray}
	\label{eq:PMeq}
	\sigma(\mathrm{p}_1(P_1) &+& \mathrm{p}_2(P_2) \rightarrow j_1(p_1) + j_2(p_2) ) \nonumber \\
	&=& \sum_i \sum_j \int^1_0d\xi_1 \int^1_0d\xi_2 \sigma(i(p_i) + j(p_j) \rightarrow k(p_k) + l(p_l))N_i(\xi_1)N_j(\xi_2),
\end{eqnarray}
 where $\mathrm{p}_1$, $\mathrm{p}_2$ denote incoming protons,
 $j_1$, $j_2$ denote observed jets,
 $\xi_1$, $\xi_2$ are the momentum fractions
\begin{eqnarray}
	\xi_1 = \frac{p_i}{P_1}, \qquad \xi_2 = \frac{p_j}{P_2},
\end{eqnarray}
 and $N_i(\xi_1)$, $N_j(\xi_2)$
 are the parton distribution functions of protons for initial partons $i$ and $j$, respectively.
The cross section in Eq.(\ref{eq:PMeq}) can be written in the form
\begin{eqnarray}
	\label{eq:PMeqtransformed}
	&{\int}&\kern-.1cm{dM^2} \,
	 \frac{d\sigma(\mathrm{p}_1(P_1) + \mathrm{p}_2(P_2) \rightarrow j_1(p_1) + j_2(p_2) )}{dM^2} \nonumber \\
	&=& {\int}\kern-.1cm{dM^2} \sum_i \sum_j {\int^1_0}\kern-.1cm{d\xi_1} {\int^1_0}\kern-.1cm{d\xi_2} \,
	 \sigma(i(p_i) + j(p_j) \rightarrow k(p_k) + l(p_l))N_i(\xi_1)N_j(\xi_2)\delta(M^2 - \hat{s}),
\end{eqnarray}
 where $M=\sqrt{\hat{s}}$ is the dijet invariant mass
 assuming that two partons result two jets.

Partons are scattered in the frame which is boosted from proton center-of-mass frame.
When a boost velocity from the parton center-of-mass frame to the proton center-of-mass frame
 is given by $\beta \equiv \tanh Y$, pseudo-rapidities of observed jets, $y_1$ and $y_2$, are written as
\begin{eqnarray}
	y_1 = Y + y, \qquad y_2 = Y - y,
\end{eqnarray} 
 where $y$ is the pseudo-rapidity of outgoing partons, $k$ and $l$, in the parton center-of-mass frame.
By taking independent variables as
\begin{eqnarray}
	\label{eq:rapidities}
	y = \frac{1}{2}(y_1 -y_2), \qquad Y = \frac{1}{2}(y_1 + y_2),
\end{eqnarray}
 and use the relations of
\begin{eqnarray}
	\hat{t} = -\frac{M^2}{2}\frac{e^{-y}}{\cosh y}&,& \qquad \hat{u} = -\frac{M^2}{2}\frac{e^{y}}{\cosh y}, \\
	\xi_1 = \frac{M}{\sqrt{s}} e^Y&,& \qquad \xi_2 = \frac{M}{\sqrt{s}} e^{-Y},
\end{eqnarray}
 we can obtain the formula of the dijet invariant mass distribution from Eq.(\ref{eq:PMeqtransformed}).
\begin{eqnarray}
&&	\frac{d\sigma(\mathrm{p}_1(P_1) + \mathrm{p}_2(P_2) \rightarrow j_1(p_1) + j_2(p_2))}{dM^2} \nonumber \\
	&=& \sum_i \sum_j \int dY \int dy \frac{1}{2\cosh^2y} f_i(\frac{M}{\sqrt{s}} e^Y) f_j(\frac{M}{\sqrt{s}} e^{-Y})
	    \frac{d\sigma(i(p_i) + j(p_j) \rightarrow k(p_k) + l(p_l))}{d\hat{t}},
\end{eqnarray}
 where $f_i(x) \equiv x N_i(x)$.
The possible range of $Y$ is
\begin{eqnarray}
 -\ln \frac{\sqrt{s}}{M} < Y < \ln \frac{\sqrt{s}}{M}
\end{eqnarray}
 with the proton center-of-mass energy $\sqrt{s}$.
The necessary differential cross sections of partons
 are described by the corresponding squared amplitudes with spin- and color-averaged factor as
\begin{eqnarray}
	\frac{d\sigma(ij\rightarrow kl)}{d\hat t} = \frac{1}{\hat s^2}\frac{1}{16\pi}\frac{1}{36}|{\cal M}(ij\rightarrow kl)|^2.
\end{eqnarray}

We show in Fig.~\ref{fig:distributions}
 the dijet invariant mass distributions at $\sqrt{s}=14$ TeV
 with the MRSTMCal parton distribution function of proton from \cite{PDF}.
We choose $N_f=6$, namely all the quarks couple to the KK gluon with the maximal coupling $\sqrt{2}g_s$,
 by which we can start from the case of wide resonance structures.
The kinematical cuts are chosen as $p_T > 30$ GeV or $500$ GeV and $|y|\leq2.5$, $2.0$ or $1.5$.
Since the mass of axigluon, which is very similar to the KK gluon in this benchmark model,
 is constrained to be heavier than 3.6 TeV by the CMS experiment at the LHC \cite{Khachatryan:2015sja},
 we take the mass of first KK gluon as 4, 5, 6 or 7 TeV.
We see in Fig.~\ref{fig:distributions} that rapidity cut is more effective than $p_T$ cut.
The stronger rapidity cuts give less number of QCD events
 with almost no reduction of the number of events by KK gluon exchange.
This is due to the fact
 that the production of a single heavy particle, namely the KK gluon, in $s$-channel
 contributes to the events with larger scattering angles.
The $p_T$ cut is relatively effective in case of weaker rapidity cuts.

\begin{figure}[t]
  \begin{center}
      \includegraphics[width=17.0cm]{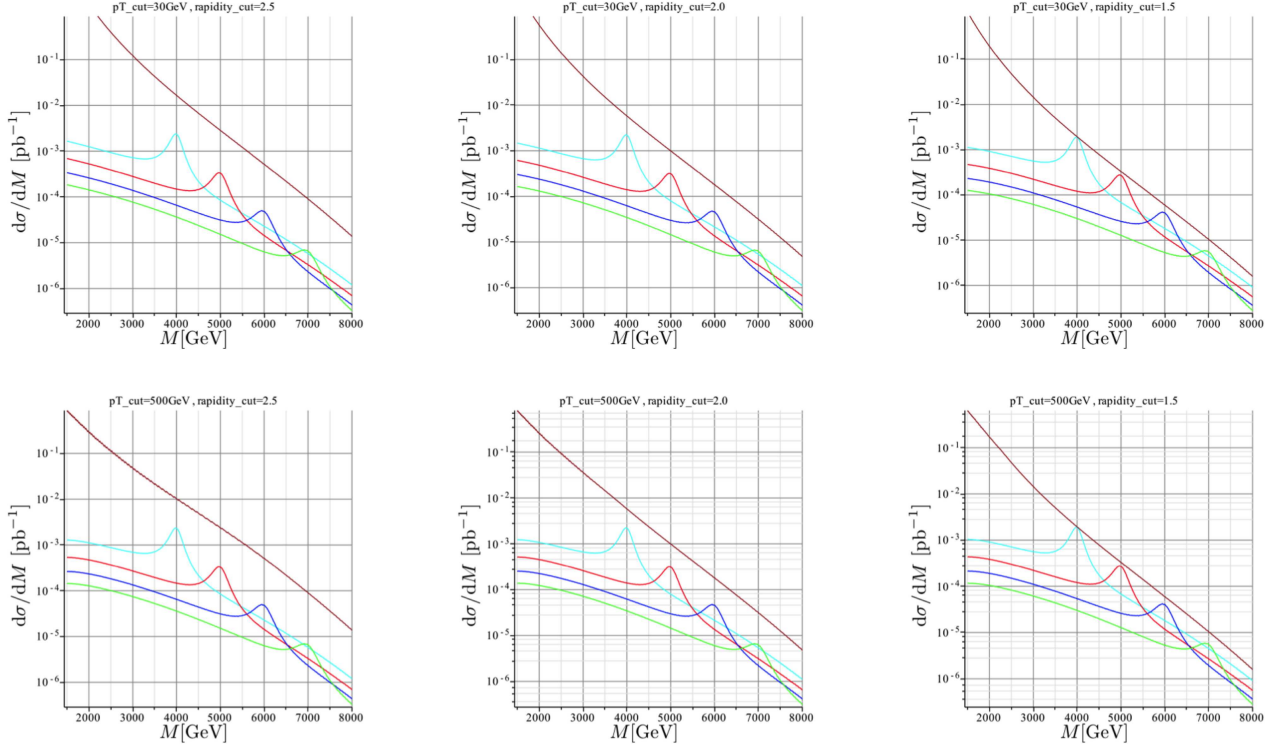}
   \caption{
Dijet invariant mass distributions
 by ordinary two body QCD scattering processes (brown) and by one KK gluon exchanges
 in case with rapidity cuts $|y|\leq$ 2.5 (left column), 2.0 (middle column) and 1.5 (right column)
 with $p^{\rm min}_T=30$ GeV (upper row) and $p^{\rm min}_T=500$ GeV (lower row)
 at $\sqrt{s}=14$ TeV.
We assume $N_f=2$ in the amplitude of one KK gluon exchanges.
The lines with cyan, red, blue and green
 correspond to the KK gluon masses $m_1=4, 5, 6$ and $7$ TeV, respectively.
} 
\label{fig:distributions-2}
 \end{center}
\end{figure}
\begin{table}[t] 
   \begin{center}
   \begin{tabular}{c}
\begin{minipage}{0.5\hsize}
 \begin{center}
   \begin{tabular}{|c||c|c|c|} 
  \multicolumn{4}{c}{$S/\sqrt{B}$ with $\int Ldt= 25\mathrm{\ fb}^{-1}$} \\ \hline\hline
  \multicolumn{1}{|c||}{maximum of $|y|$} & 2.5 & 2.0 & 1.5  \\ \hline\hline
  \multicolumn{4}{|c|}{$p_T^{\rm min}$ = $30$ GeV}  \\ \hline
  $m_1$ = 4 TeV  & 35 & 55 & 81 \\
  $m_1$ = 5 TeV  & 12 & 20 & 29 \\
  $m_1$ = 6 TeV  & 4.2 & 6.8 & 10 \\ 
  $m_1$ = 7 TeV  & 1.4 & 2.2 & 3.4 \\ \hline\hline
    \multicolumn{4}{|c|}{$p_T^{\rm min}$ = $500$  GeV} \\ \hline
  $m_1$ = 4 TeV  & 44 & 55 & 81 \\
  $m_1$ = 5 TeV  & 13 & 20 & 29 \\
  $m_1$ = 6 TeV  & 4.2 & 6.8 & 10 \\ 
  $m_1$ = 7 TeV  & 1.4 & 2.2 & 3.4 \\ \hline
  \end{tabular}
 \end{center}
\end{minipage}
    \begin{minipage}{0.5\hsize}
 \begin{center}
   \begin{tabular}{|c||c|c|c|} 
  \multicolumn{4}{c}{$S/\sqrt{B}$ with $\int Ldt= 100\mathrm{\ fb}^{-1}$}\\ \hline\hline
  \multicolumn{1}{|c||}{maximum of $|y|$} & 2.5 & 2.0 & 1.5  \\ \hline\hline
  \multicolumn{4}{|c|}{$p_T^{\rm min}$ = $30$ GeV} \\ \hline
  $m_1$ = 4 TeV  & 70 & 110 & 160 \\
  $m_1$ = 5 TeV  & 24 & 39 & 59 \\
  $m_1$ = 6 TeV  & 8.4 & 14 & 21 \\ 
  $m_1$ = 7 TeV  & 2.8 & 4.5 & 6.9 \\ \hline\hline
    \multicolumn{4}{|c|}{$p_T^{\rm min}$ = $500$ GeV} \\ \hline
  $m_1$ = 4 TeV  & 87 & 110 & 160 \\
  $m_1$ = 5 TeV  & 26 & 39 & 59 \\
  $m_1$ = 6 TeV  & 8.5 & 14 & 21 \\ 
  $m_1$ = 7 TeV  & 2.8 & 4.5 & 6.9 \\ \hline
  \end{tabular}
 \end{center}
\end{minipage}
\end{tabular}
\end{center}
\caption{
Signal-to-noise ratios
 corresponding to various masses of KK gluon and kinematical cuts
 in case of $N_f=2$ at $\sqrt{s} = 14$ TeV
 with the integrated luminosity $\int Ldt= 25\mathrm{\ fb}^{-1}$ (Left table)
 and $\int Ldt= 100\mathrm{\ fb}^{-1}$ (Right table).
}
\label{tab:SNratio-2}
\end{table}

We employ the signal-to-noise ratio to present the statistical significance of the signals.
We define the signal-to-noise ratio
 as the number of signal events by KK gluon exchanges
 divided by square root of the number of events by QCD processes.
The number of events is obtained
 by integrating the dijet invariant mass distribution
 over the range from $m_1-250$ GeV to $m_1+250$ GeV independent from the value of $m_1$.
We consider only the events near the resonance by the KK gluon,
 because we are neglecting the interference effect between KK gluon processes and QCD processes.
The signal-to-noise ratios
 corresponding to 25 $\mathrm{fb}^{-1}$ and 100 $\mathrm{fb}^{-1}$ of integrated luminosity at $\sqrt{s} = 14$ TeV
 are given in Table \ref{tab:SNratio}.
If we take the criteria of discovery
 as that the value of signal-to-noise ratio is larger than 5,
 the LHC with $\sqrt{s} = 14$ TeV and luminosity 100 fb$^{-1}$ has ability to detect $m_1 \lesssim 6.2$ TeV
 with $|y|\leq2.5$ and $m_1 \lesssim 7.2$ TeV with $|y|\leq1.5$.
The number of signal events is 127
 when KK gluon mass is 7 TeV at $|y|\leq 1.5$ with the integrated luminosity of 100 $\mathrm{fb}^{-1}$.

Now we consider the model dependence.
We first reduce the number of quarks with the maximal coupling from $N_f=6$ to $N_f=2$,
 namely, only up and down quarks can couple with the KK gluon.
We show in Fig.~\ref{fig:distributions-2} dijet invariant mass distributions in this case.
The structure of the resonance becomes more evident due to the reduction of the decay width.
The integration range of the dijet invariant mass distribution to obtain the number of events
 is now chosen from $m_1 - 80$ GeV to $m_1 + 80$ GeV independent from the value of $m_1$
 corresponding to this reduction of the width.
We always conservatively choose the range 
 so that it coincides with the width of the KK gluon of $m_1=2.5$ TeV.
Table \ref{tab:SNratio-2} shows signal-to-noise ratios
 in case of various masses of the KK gluon and kinematical cuts.
The significance of the signal is increased in comparison with the case of $N_f=6$
 due to the effect of narrower resonance, or smaller width of the KK gluon.
Though the number of signal events are almost independent from $N_f$,
 the number of background QCD events behaves roughly as $N_f$
 due to the change of integration range.
Therefore, the signal-to-noise ratio roughly depends on the inverse of the square root of $N_f$.
If we take the criteria of discovery
 as that the value of signal-to-noise ratio is larger than 5,
 the LHC with $\sqrt{s} = 14$ TeV and luminosity 100 fb$^{-1}$ has ability to detect $m_1 \lesssim 6.5$ TeV
 with $|y|\leq2.5$ and $m_1 \lesssim 7.4$ TeV with $|y|\leq1.5$.
The number of signal events is 92
 when KK gluon mass is 7 TeV at $|y|\leq 1.5$ with the integrated luminosity of 100 $\mathrm{fb}^{-1}$.

\begin{table}[h] 
   \begin{center}
   \begin{tabular}{c}
\begin{minipage}{0.5\hsize}
 \begin{center}
   \begin{tabular}{|c||c|c|c|} 
  \multicolumn{4}{c}{Reaches for $f$ with $\int Ldt= 25\mathrm{\ fb}^{-1}$} \\ \hline\hline
  \multicolumn{1}{|c||}{maximum of $|y|$} & 2.5 & 2.0 & 1.5  \\ \hline\hline
  \multicolumn{4}{|c|}{$p_T^{\rm min}$ = $30$ GeV}  \\ \hline
  $m_1$ = 4 TeV  & 0.62 & 0.55 & 0.50 \\
  $m_1$ = 5 TeV  & 0.80 & 0.71 & 0.64 \\
  $m_1$ = 6 TeV  & - & 0.93 & 0.83 \\ 
  $m_1$ = 7 TeV  & - & - & - \\ \hline\hline
    \multicolumn{4}{|c|}{$p_T^{\rm min}$ = $500$  GeV} \\ \hline
  $m_1$ = 4 TeV  & 0.58 & 0.55 & 0.50 \\
  $m_1$ = 5 TeV  & 0.78 & 0.71 & 0.64 \\
  $m_1$ = 6 TeV  & - & 0.93 & 0.83 \\ 
  $m_1$ = 7 TeV  & - & - & - \\ \hline
  \end{tabular}
 \end{center}
\end{minipage}
    \begin{minipage}{0.5\hsize}
 \begin{center}
   \begin{tabular}{|c||c|c|c|} 
  \multicolumn{4}{c}{Reaches for $f$ with $\int Ldt= 100\mathrm{\ fb}^{-1}$} \\ \hline\hline
  \multicolumn{1}{|c||}{maximum of $|y|$} & 2.5 & 2.0 & 1.5  \\ \hline\hline
  \multicolumn{4}{|c|}{$p_T^{\rm min}$ = $30$ GeV} \\ \hline
  $m_1$ = 4 TeV  & 0.52 & 0.46 & 0.42 \\
  $m_1$ = 5 TeV  & 0.67 & 0.60 & 0.54 \\
  $m_1$ = 6 TeV  & 0.88 & 0.78 & 0.70 \\ 
  $m_1$ = 7 TeV  & - & - & 0.92 \\ \hline\hline
    \multicolumn{4}{|c|}{$p_T^{\rm min}$ = $500$ GeV} \\ \hline
  $m_1$ = 4 TeV  & 0.49 & 0.46 & 0.42 \\
  $m_1$ = 5 TeV  & 0.66 & 0.60 & 0.54 \\
  $m_1$ = 6 TeV  & 0.88 & 0.78 & 0.70 \\ 
  $m_1$ = 7 TeV  & - & - & 0.92 \\ \hline
  \end{tabular}
 \end{center}
\end{minipage}
\end{tabular}
\end{center}
\caption{
Minimum accessible values of the parameter $f ({} \leq 1)$
 corresponding to $5\sigma$ significance with various kinematical cuts at $\sqrt{s} = 14$ TeV.
Left and right tables correspond the case with the integrated luminosity $\int Ldt= 25\mathrm{\ fb}^{-1}$
 and $\int Ldt= 100\mathrm{\ fb}^{-1}$, respectively.
}
\label{tab:coefficient-f}
\end{table}

It has been shown that
 the coupling between quarks and KK gluon depends on the physics of Yukawa coupling generation.
Though our benchmark model gives the relation $g^{(1)}_s=\sqrt{2}g_s$ for light quarks (up and down quarks),
 it is not the model independent universal result.
We introduce a new parameter $f$
 by replacing the coupling $\sqrt{2}g_s$ with $f \sqrt{2}g_s$ in previous formulae in case of $N_f=2$.
It is easy to see that
 the value of $f$ is less than $1$ with general anti-symmetric functions
 instead of $\epsilon(y)$ in Eq.(\ref{eq:toylagrangian}).
Therefore, 
 we investigate the result with a conservative range of $0<f<1$.
It is easy to translate the signal-to-noise ratio to any model with $f \neq 1$.
We may simply multiply $f^4$ to the signal-to-noise ratio in case of $f=1$.
We do not include $f$ in the formula of decay width
 to avoid its underestimate for small value of $f$
 by complete disregard of the other decay channels to heavier quarks.
The observable effect of 6 TeV first KK gluon with the significance over 5$\sigma$
 is appeared in a case $f \gtrsim 0.78$
 for the integrated luminosity of 100 fb$^{-1}$
 with the kinematical cut $|y| \leq 2.0$ and $p_T^{\rm min} = 500$ GeV, for example.
We show in Table \ref{tab:coefficient-f}
 the minimum accessible values of $f$ in various cases of KK gluon masses and kinematical cuts.
  
The KK gluon as an axial vector particle is also a special result in our benchmark model,
 which also depends on the form of the bulk mass, or depends on the physics of Yukawa coupling generation.
If the KK gluon is not axial vector,
 its single production by gluon fusion through the triangle anomaly diagram may be significant,
 which have not been included in the above analysis.
However, it has been shown that
 such a diagram vanishes in case of no chiral anomaly in QCD color current \cite{ArkaniHamed:2001is}.
Therefore, we expect that
 gluon fusion does not contribute to produce KK gluons in GHU models.
Even though it could contribute,
 it has been shown that the effects are negligibly small by general analysis \cite{Chivukula:2013xla}.
In this way we can translate the results in our benchmark model to those in any GHU models with KK gluons.

\section{Summary}
\label{sec:summary}

We have examined the possibility of observing
 the first KK gluon in GHU models at the LHC experiments with $\sqrt{s} =14$ TeV.
We took a benchmark model with gauge symmetry SU$(3)_C \times$ SU$(3)_W$ in five-dimensional space-time.
It is natural that
 the gauge symmetry of the strong interaction is introduced in the bulk
 as well as the gauge symmetry which reduces to the electroweak gauge symmetry.
The fifth dimension was compactified on $S^1/{\bf Z}_2$ orbifold.
Quarks, especially up and down quarks, were introduced as bulk fields with bulk masses
 so that their non-trivial configurations in fifth dimension
 could generate appropriate Yukawa couplings with the Higgs doublet field.
Though the couplings of the first KK gluon to up and down quarks depend on bulk masses,
 we argue that the values of the couplings are the same $\sqrt{2}g_s$ in good approximation,
 since the Yukawa couplings of up and down quarks are very small.
It is worth to mention that
 in the models in which quarks are localized at orbifold fixed points,
 it has been shown that the couplings of KK gluon with massless quarks are also $\sqrt{2}g_s$ \cite{Dicus:2000hm}.
In our benchmark model we assumed that
 the couplings to the other four quarks were also the same value,
 by which we could start from wide resonance structures in dijet invariant mass distributions.
The contribution of the first KK gluon in this benchmark model
 to the dijet invariant mass distributions at the LHC was investigated
 with various rapidity cuts and $p_T$ cuts in the parton level.
In cases of the integrated luminosity of 25 $\mathrm{fb}^{-1}$ and 100 $\mathrm{fb}^{-1}$
 we estimated signal-to-noise ratios, $S/\sqrt{B}$,
 with kinematical cuts of jet rapidity $|y|\leq2.5, 2.0$ and $1.5$ and jet $p_T\geq30$ and $500$ GeV.
We found in this benchmark model that
 in case of the integrated luminosity of 100 $\mathrm{fb}^{-1}$
 the first KK gluon mass up to $6.2$ TeV and $7.2$ TeV could be excluded at $5\sigma$ level
 with rapidity cuts $|y| \leq 2.5$ and $|y| \leq 1.5$, respectively.

The dependence of the results on the models of GHU was examined first by assuming that
 the couplings of four heavier quarks than up and down quarks did not couple with the first KK gluon.
In case of the integrated luminosity of 100 $\mathrm{fb}^{-1}$
 the first KK gluon with mass up to $6.5$ TeV and $7.4$ TeV could be excluded at $5\sigma$ level
 with rapidity cuts $|y| \leq 2.5$ and $|y| \leq 1.5$, respectively.
We further introduced a parameter $f$
 in order to apply the results in the model with $N_f=2$ to general GHU models with KK gluons.
The coupling between the first KK gluon with up and down quarks, $\sqrt{2}g_s$,
 was replaced by $f\sqrt{2}g_s$,
 since it could depend on the physics of Yukawa coupling generation.
The signal-to-noise ratios in the benchmark model
 were easily translated to those of the general models by multiplying the factor $f^4$.
We have investigated the minimum accessible values of $f$
 corresponding to $5\sigma$ significance with various KK gluon masses and kinematical cuts.
We found, for example, that
 $f \geq 0.78$ is necessary to observe the signal of 6 TeV first KK gluon
 with integrated luminosity of 100 fb$^{-1}$ under $|y| \leq 2.0$ and $p_T \geq 500$ GeV.

More detailed analysis beyond the parton level requires the techniques of Monte Carlo simulations.
It could be possible that
 the constraint to the size of the compact space would be stronger
  than that was obtained from the effects of KK weak bosons.
It would be interesting to clarify
 whether or not the obtained constraint contradicts to the observed electroweak scale
 in concrete promising GHU models. 
It is also important to investigate the strategy
 to distinguish the signal of KK gluons of GHU models from similar in other models.
Investigations into the width,
 the angular distributions of dijets
 and searching for the second KK gluon
 are possible strategies.
We leave these issues to future research.

\section*{Acknowledgements}
We would like to thank C.S.~Lim, S.~Kohara and S.~Fukasawa for helpful comments and discussions.
N.K. was supported in part by JSPS KAKENHI Grant Number 26400253.
Y.S. was supported in part by JGC-S SCHOLARSHIP FOUNDATION.



\begin{thebibliography}{99}

\bibitem{Antoniadis:1990ew}
  I.~Antoniadis,
  Phys.\ Lett.\ B {\bf 246} (1990) 377.

\bibitem{ArkaniHamed:1998rs}
  N.~Arkani-Hamed, S.~Dimopoulos and G.~R.~Dvali,
  Phys.\ Lett.\ B {\bf 429} (1998) 263.

\bibitem{ArkaniHamed:1998nn}
  N.~Arkani-Hamed, S.~Dimopoulos and G.~R.~Dvali,
  Phys.\ Rev.\ D {\bf 59} (1999) 086004.
  
\bibitem{Kawamura:1999nj}
  Y.~Kawamura,
  Prog.\ Theor.\ Phys.\  {\bf 103} (2000) 613.

\bibitem{Manton:1979kb}
  N.~S.~Manton,
  Nucl.\ Phys.\ B {\bf 158} (1979) 141.

\bibitem{Fairlie:1979at}
  D.~B.~Fairlie,
  Phys.\ Lett.\ B {\bf 82} (1979) 97.

\bibitem{Hosotani:1983xw}
  Y.~Hosotani,
  Phys.\ Lett.\ B {\bf 126} (1983) 309.

\bibitem{Hosotani:1983vn}
  Y.~Hosotani,
  Phys.\ Lett.\ B {\bf 129} (1983) 193.

\bibitem{Hosotani:1988bm}
  Y.~Hosotani,
  Annals Phys.\  {\bf 190} (1989) 233.

\bibitem{Hatanaka:1998yp}
  H.~Hatanaka, T.~Inami and C.~S.~Lim,
  Mod.\ Phys.\ Lett.\ A {\bf 13} (1998) 2601.

\bibitem{Antoniadis:2001cv}
  I.~Antoniadis, K.~Benakli and M.~Quiros,
  New J.\ Phys.\  {\bf 3} (2001) 20.

\bibitem{Hall:2001zb}
  L.~J.~Hall, Y.~Nomura and D.~Smith,
  Nucl.\ Phys.\ B {\bf 639} (2002) 307.

\bibitem{Csaki:2003dt}
  C.~Csaki, C.~Grojean, H.~Murayama, L.~Pilo and J.~Terning,
  Phys.\ Rev.\ D {\bf 69} (2004) 055006.

\bibitem{Burdman:2002se}
  G.~Burdman and Y.~Nomura,
  Nucl.\ Phys.\ B {\bf 656} (2003) 3.

\bibitem{Scrucca:2003ra}
  C.~A.~Scrucca, M.~Serone and L.~Silvestrini,
  Nucl.\ Phys.\ B {\bf 669} (2003) 128.

\bibitem{Maru:2007xn}
  N.~Maru and N.~Okada,
  Phys.\ Rev.\ D {\bf 77} (2008) 055010.

\bibitem{Maru:2013ooa}
  N.~Maru and N.~Okada,
  Phys.\ Rev.\ D {\bf 87} (2013) 9,  095019.

\bibitem{Maru:2013bja}
  N.~Maru and N.~Okada,
  Phys.\ Rev.\ D {\bf 88} (2013) 3,  037701.

\bibitem{Nath:1999mw}
  P.~Nath, Y.~Yamada and M.~Yamaguchi,
  Phys.\ Lett.\ B {\bf 466} (1999) 100.

\bibitem{Dicus:2000hm}
  D.~A.~Dicus, C.~D.~McMullen and S.~Nandi,
  Phys.\ Rev.\ D {\bf 65} (2002) 076007.

\bibitem{Agashe:2006hk}
  K.~Agashe, A.~Belyaev, T.~Krupovnickas, G.~Perez and J.~Virzi,
  Phys.\ Rev.\ D {\bf 77} (2008) 015003.

\bibitem{Lillie:2007ve}
  B.~Lillie, J.~Shu and T.~M.~P.~Tait,
  Phys.\ Rev.\ D {\bf 76} (2007) 115016.

\bibitem{Haba:2012bc}
  N.~Haba, K.~Kaneta and S.~Tsuno,
  Phys.\ Rev.\ D {\bf 87} (2013) 9,  095002.

\bibitem{Chivukula:2011ng}
  R.~S.~Chivukula, A.~Farzinnia, E.~H.~Simmons and R.~Foadi,
  Phys.\ Rev.\ D {\bf 85} (2012) 054005.

\bibitem{Chivukula:2013xla}
  R.~S.~Chivukula, A.~Farzinnia, J.~Ren and E.~H.~Simmons,
  Phys.\ Rev.\ D {\bf 87} (2013) 9,  094011.

\bibitem{Frampton:1987dn}
  P.~H.~Frampton and S.~L.~Glashow,
  Phys.\ Lett.\ B {\bf 190} (1987) 157.

\bibitem{PDF}
 The Durham HepData Project, http://hepdata.cedar.ac.uk/pdf/pdf3.html

\bibitem{Khachatryan:2015sja}
  V.~Khachatryan {\it et al.} [CMS Collaboration],
  Phys.\ Rev.\ D {\bf 91} (2015) 5,  052009.
  
\bibitem{ArkaniHamed:2001is}
  N.~Arkani-Hamed, A.~G.~Cohen and H.~Georgi,
  Phys.\ Lett.\ B {\bf 516} (2001) 395.




\end{thebibliography}
\end{document}